\documentclass[fleqn,twoside]{article}
\usepackage{espcrc2}

\usepackage{graphicx}
\usepackage[figuresright]{rotating}

\newcommand{\AmS}{{\protect\the\textfont2
  A\kern-.1667em\lower.5ex\hbox{M}\kern-.125emS}}

\title{A Phenomenological Treatment of Chiral Symmetry Restoration and 
Deconfinement}
 
\author{Peter N. Meisinger
        \address[stupidaddressmark] {Dept. of Physics,
        Washington University, \\
        St. Louis, MO 63130 USA}%
        \thanks{We gratefully acknowledge the support of the U.S. Dept. of
                Energy under DOE DE-FG02-91ER40628},
        Travis R. Miller \addressmark,
        and Michael C. Ogilvie \addressmark[stupidaddressmark]} 

\begin{document}

\begin{abstract}
A phenomenological expression for the thermodynamic potential of gluons
and quarks is constructed which incorporates the features of deconfinement
and chiral symmetry restoration known from lattice simulations. The
thermodynamic potential is a function of the Polyakov loop and chiral
condensate expectation values. The gluonic sector uses a successful
model for pure \( SU(N_{c}) \) gauge theories in which the Polyakov
loop eigenvalues are the fundamental order parameters for deconfinement.
The quark sector is given by a Nambu-Jona-Lasinio model in which a
constant background \( A_{0} \) field couples the chiral condensate
to the Polyakov loop. We consider the case of \( N_{f}=2 \) in detail.
For two massless quarks, we find a second order chiral phase transition.
Confinement effects push the transition to higher temperatures, but
the entropy associated with the gluonic sector acts in the opposite
direction. For light mass quarks, only a rapid crossover occurs. For
sufficiently heavy quarks, a first order deconfinement transition
emerges. This simplest model has one adjustable parameter, which can
be set from the chiral transition temperature for light quarks. It
predicts all thermodynamic quantities as well as the behavior of the
chiral condensate and the Polyakov loop over a wide range of temperatures.
\vspace{1pc}
\end{abstract}

\maketitle

Finite temperature QCD, in its variants which can be studied with
lattice techniques, exhibits both deconfinment and chiral symmetry
restoration. Depending on the model under study, there may be one
or more phase transitions as the temperature is varied, or no phase
transition at all. Finite density brings with it an additional rich
set of possible behaviors at low temperatures. The transition from
one phase to another may be marked by a first- or second-order phase
transition, or two apparently different phases may be connected to
one another, as in a liquid-gas transition. While we understand much
about these complex behaviors, we lack a comprehensive, simple theoretical
tool which gives us insight into possible behaviors. In short, we
need a Landau-Ginzburg theory for finite temperature QCD.

The behavior of finite temperature QCD can be understood from the
behavior of two order parameters: the Polyakov loop \( P \), associated
with deconfinement, and the chiral condensate, \( \left\langle
\bar{\psi }\psi \right\rangle  \),
associated with chiral symmetry restoration. 
Our aim is to construct simple phenomenological models incorporating
both chiral symmetry restoration and deconfinement in a unified way,
which can be applied to both quenched and unquenched behavior, for
all values of \( N_{c} \) and any number of flavors \( N_{f} \).
These models naturally give the behavior of \( P \), \( \left\langle 
\bar{\psi }\psi \right\rangle  \),
and thermodynamic variable such as pressure as a function of temperature
and chemical potential. By incorporating both \( P \) and \( \left\langle 
\bar{\psi }\psi \right\rangle  \)
as order parameters, useful information about non-equilibrium behavior
is also available.

Polyakov loop models have been used extensively to describe the deconfinement
transition in pure gauge theories
\cite{Svetitsky:1982gs}\cite{Pisarski:2000eq}.
The two models of the deconfinement transition in \( SU(N_{c}) \)
gauge theories which we have developed
\cite{Meisinger:2001cq}.
use the eigenvalues of
the Polyakov loop as the fundamental order parameters. This accords
with the results of high-temperature perturbation theory
\cite{Meisinger:2001fi},
and our models reproduce the leading \( T^{4} \) behavior of perturbation
theory at high temperature. 
Both of the models we have studied correctly predict a second
order phase transition for \( SU(2) \) and a first order phase transition
for \( SU(N_{c}) \) when \( N_{c}\geq 3 \). 

In the case of \( SU(3) \), the center symmetry \( Z(3) \) allows
us to consider the vacuum expectation value of the trace of the Polyakov
loop to lie along the real axis. The trace of the Polyakov loop in
the fundamental representation can be parametrized along the real
axis. as \( P_{F}=1+\cos (2\pi /3-\phi ) \), where \( \phi =0 \)
corresponds to the confined phase and \( \phi \neq 0 \) corresponds
to a non-zero value for \( P_{F} \) signaling deconfinement. We will
use our first model, model A, to describe deconfinement here. The
free energy as a function of \( \phi  \) has the form
\begin{eqnarray}
V_{G}=\frac{8\pi ^{2}T^{4}}{405}+\left( \frac{3T^{2}M^{2}}{2\pi ^{2}}
-\frac{2T^{4}}{3}\right) \phi ^{2} \nonumber \\
-\frac{2T^{4}}{3\pi }\phi ^{3}-\frac{3T^{4}}{2\pi ^{2}}\phi ^{4}.
\end{eqnarray}
This model is obtained from a truncation of the free energy of massive
gluons moving in a constant Polyakov loop background. Model A has
a single free parameter, the mass scale \( M \).
The pressure, energy density
and interaction measure derived from the above thermodynamic potential
compare well with pure gauge simulations.

Quarks break \( Z(3) \) symmetry explicitly. This effect is small
for large quark mass, and the deconfinement transition remains first
order for sufficiently large quark masses. Within model A, we find
that for a single heavy quark, deconfinement persists as a first-order
phase transition until a second-order critical point is reached at
\( m_{c}\simeq 2.57M \). For quark masses below this value, deconfinement
occurs as a smooth crossover.

A complete discussion of QCD at finite temperature must also incorporate
chiral symmetry restoration. We include these effects via a Nambu-Jona-Lasinio
(NJL) model coupled to the background Polyakov loop. The
free energy of \( N_{f}=2 \) quarks at 1-loop order can be written
as
\begin{equation}
V_{F}=\frac{\sigma ^{2}}{4G}-N_{f}N_{c}trln(i\gamma ^{\mu }D_{\mu 
}+m_{0}+\sigma ),
\end{equation}
where \( G \) is the four-fermion coupling constant, the composite
field \( \sigma =\frac{\bar{\psi }\psi }{2G} \), and \( D_{\mu } \)
is a covariant derivative containing the background \( A_{0} \) field.
The constituent mass \( m \) of the quarks is given as \( m=m_{0}+\sigma  \),
where \( m_{0} \) is the current mass. The coupling constant \( G \)
as well as the non-covariant cutoff \( \Lambda  \) required to regularize
the theory are fixed by zero temperature phenomenology. The values
used are \cite{Klevansky:qe}
\( G=5.02 \) GeV\( ^{-2} \) and \( \Lambda =653 
\)
MeV, which lead to \( f_{\pi }= 93 MeV \) 
and \( \bar{\psi }\psi  = -2(250 MeV)^3 \).
For sufficiently large temperature, the finite temperature contribution
of the functional determinant can be evaluated using a high temperature
expansion \cite{Meisinger:2001fi}.
The quark part of our model has
no remaining free parameters once it is calibrated to give the correct
zero temperature physics.

\begin{figure}
\vspace{0.1in}
\includegraphics[width=2.9in]{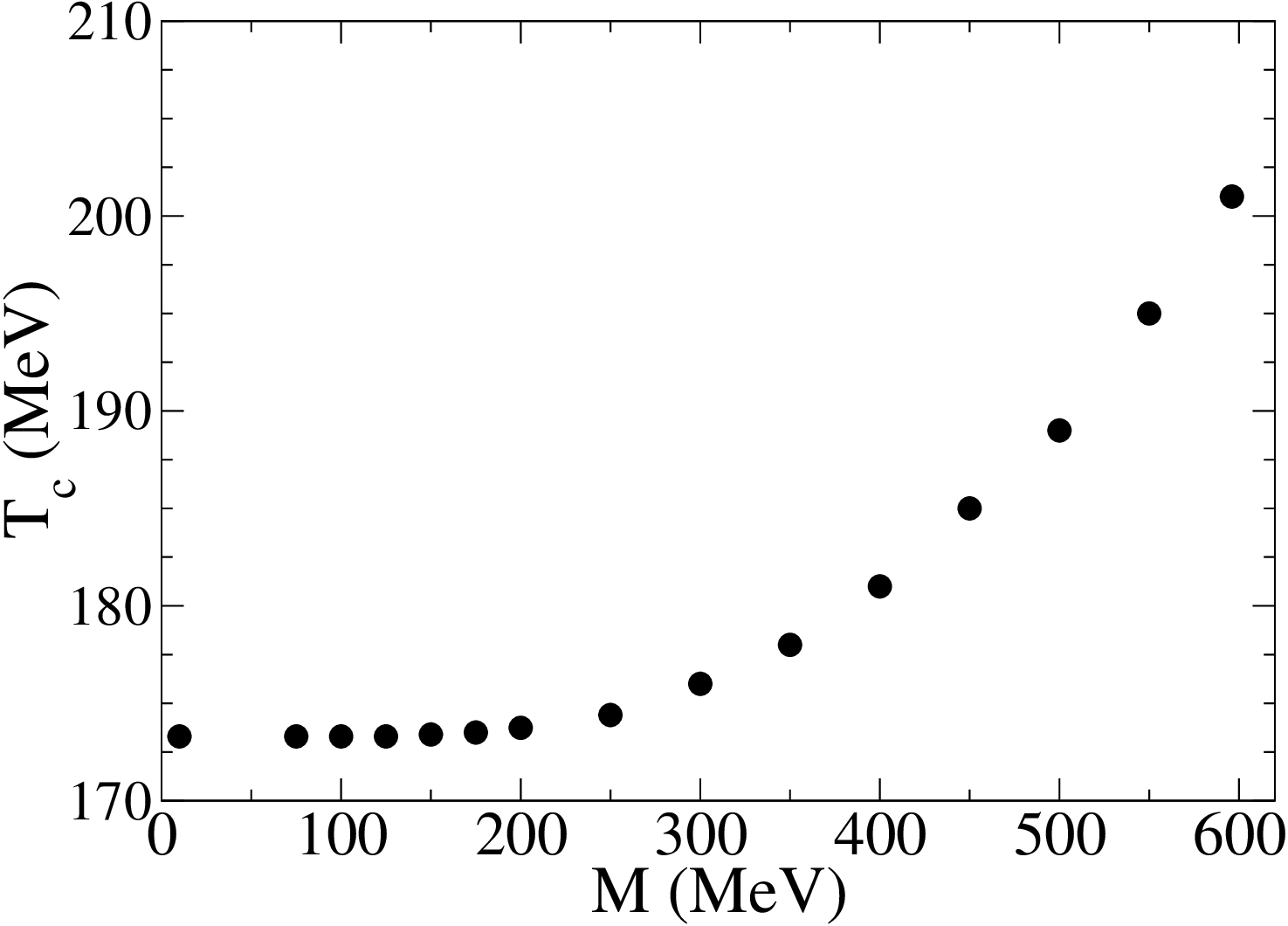}
\vspace{-0.25in}
\caption{Critical temperature $T_c$ versus $M$.}
\label{Td_vs_M}
\vspace{-0.25in}
\end{figure}

\begin{figure}
\vspace{0.3in}
\includegraphics[width=2.9in]{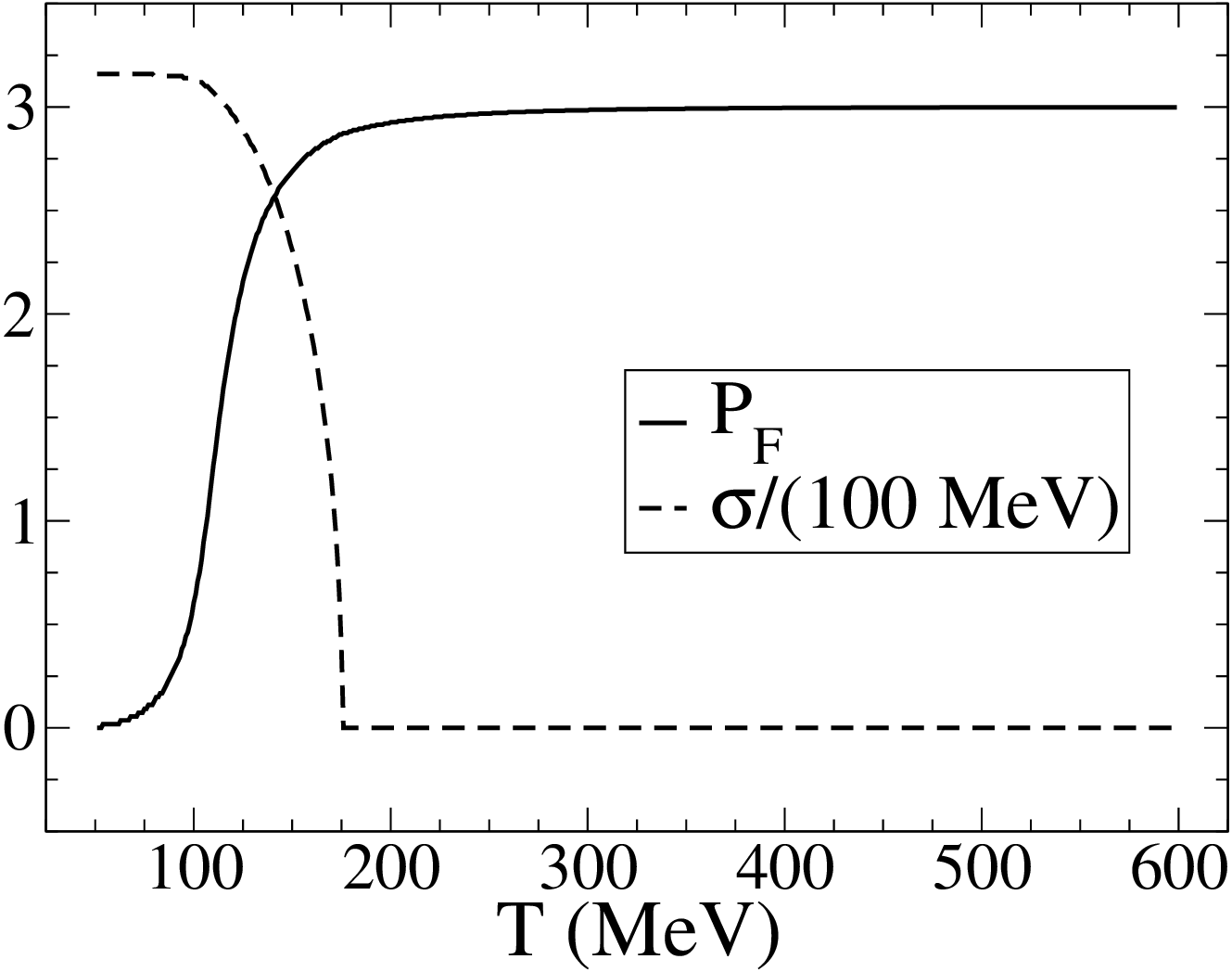}
\vspace{-0.25in}
\caption{$P$ and $\sigma / 100 \, MeV$ versus $T$ in $MeV$.}
\label{order parameters}
\vspace{-0.25in}
\end{figure}

The only free parameter in model A coupled to the NJL model is \( M \).
The chiral transition temperature \( T_{c} \) is shown in
figure \ref{Td_vs_M} as a function of \( M \). 
As M is lowered, \( T_{c} \)
decreases until  \( M=300 \)
MeV, below which \( T_{c} \) is essentially frozen at \( 175 \)
MeV, the value predicted by the NJL model with a trivial Polyakov
loop. 
At very low values of \( M \), the energy density divided
by \( T^{4} \) changes from a monotonically increasing function of
temperature to one which approaches the blackbody behavior from above
in stark constrast to lattice data. 
If one chooses \( M=596 \) MeV, the value which yields a pure
gauge deconfinement temperature of \( T_{d}=270 \) MeV, then the
chiral transition temperature is \( 201 \) MeV which is somewhat
higher than lattice simulations indicate. 
Choosing the value \( M=300 \)
MeV yields a reasonable value of the transition temperature while
avoiding the problem of poorly behaved thermodynamic functions. Once
\( M \) is fixed there are no free parameters in the model.

\begin{figure}
\vspace{0.1in}
\includegraphics[width=2.9in]{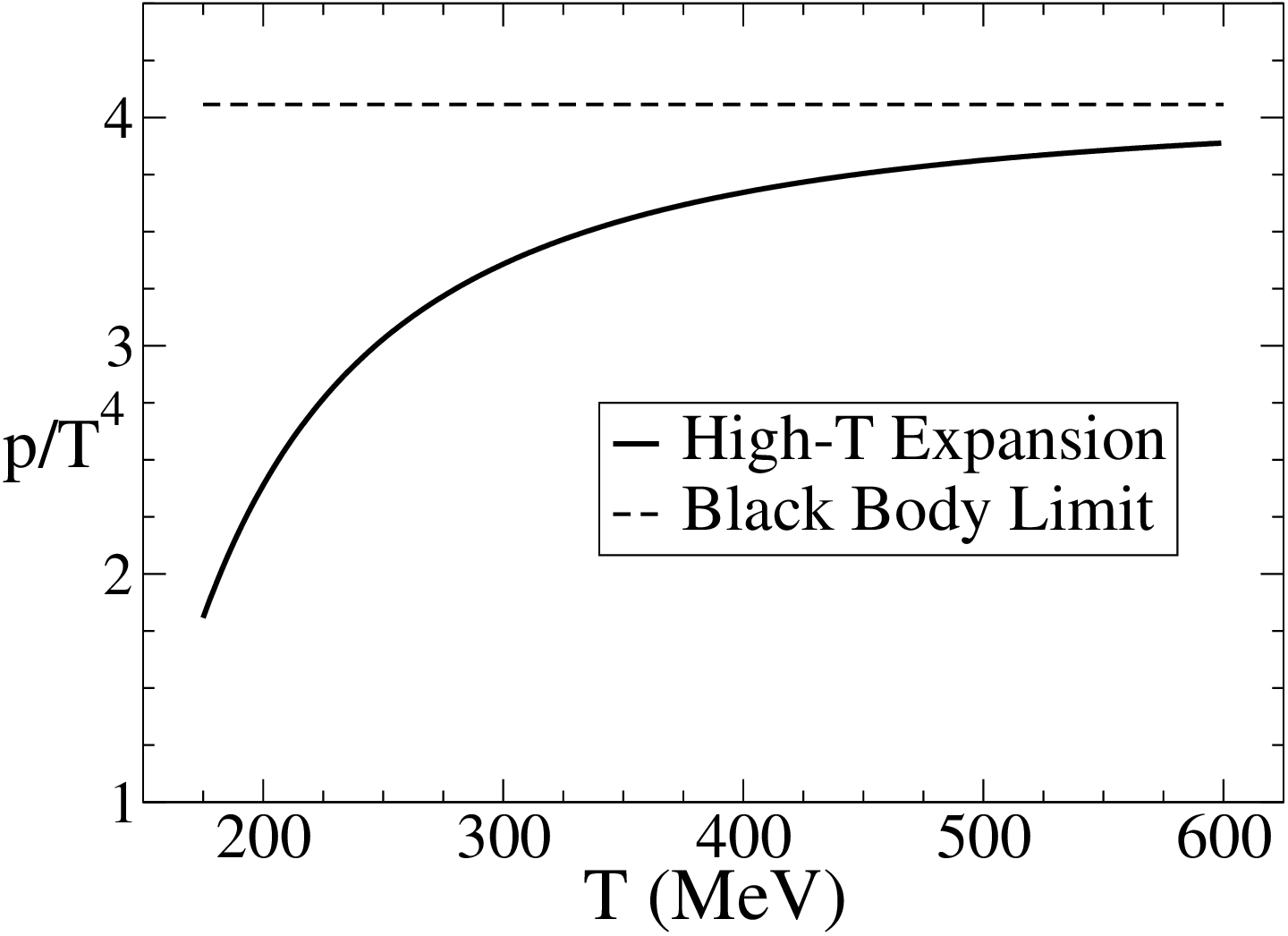}
\vspace{-0.25in}
\caption{$p/T^4$ versus $T$.}
\label{pressure}
\vspace{-0.3in}
\end{figure}

The chiral order parameter \( \sigma  \) and the Polyakov loop are
plotted as a function of temperature in figure \ref{order parameters}.
Choosing a zero current quark
mass yields a clear second order chiral transition while the
deconfinement order parameter displays crossover behavior.
The pressure divided
by \( T^{4} \) is plotted in figure \ref{pressure}.
The interaction measure $\Delta = (\epsilon - 3 p)/T^4$ falls to zero at
high temperatures as $1/T^2$, a behavior also seen in the pure gauge theory
model.

\begin{figure}
\vspace{0.1in}
\includegraphics[width=2.9in]{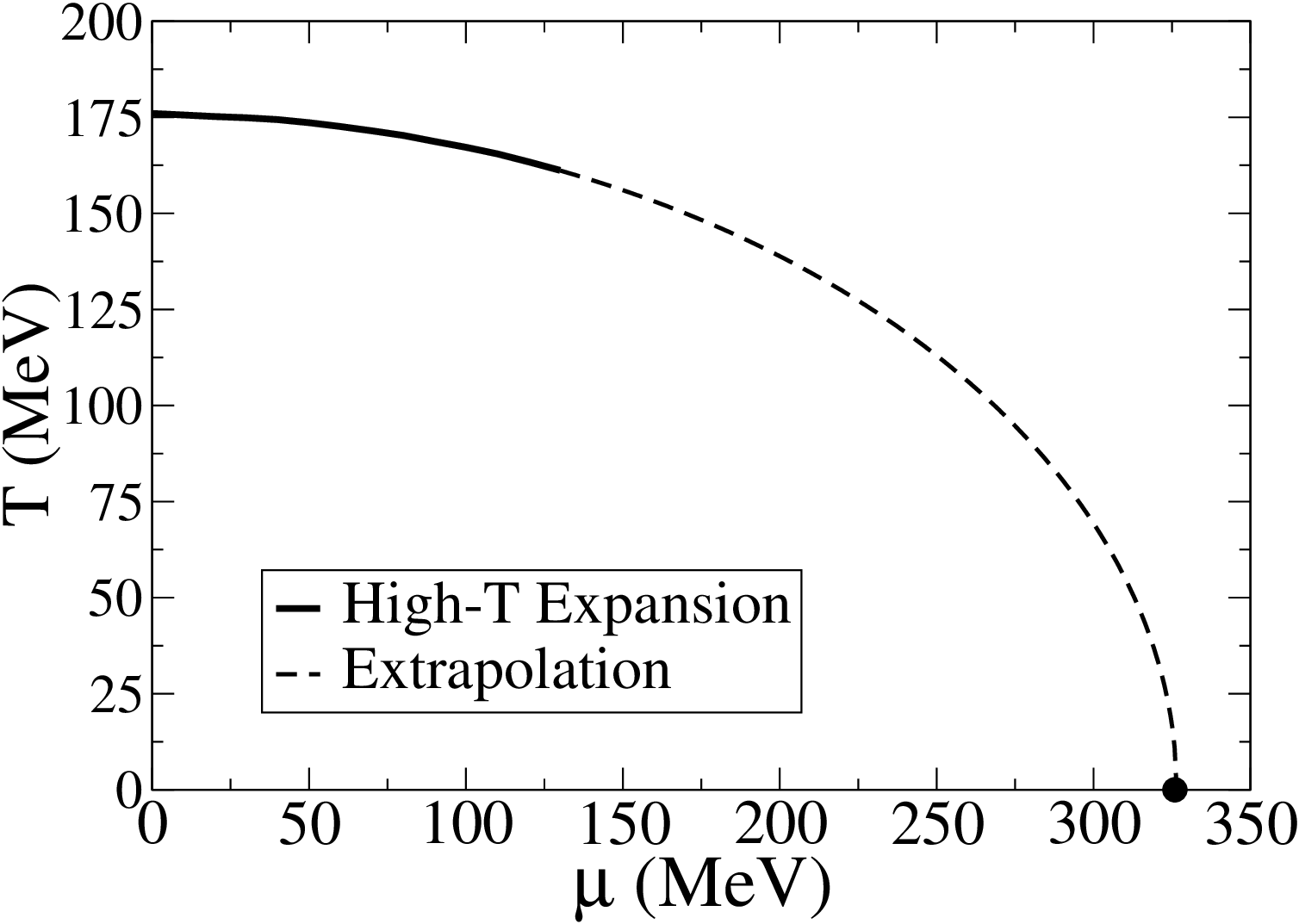}
\vspace{-0.25in}
\caption{$T_c$ versus $\mu$.}
\label{T-mu}
\vspace{-0.265in}
\end{figure}

The effect of adding a small chemical potential can also be studied
using a high temperature expansion of the quark determinant.
In figure \ref{T-mu}
the phase diagram is displayed for \( M=300 \) MeV.
The chiral transition appears to be second order out to at least
$\mu = 130 \, MeV$ which
is roughly the limit of reliability of the high-temperature expansion.
The dotted line shows an elliptical extrapolation to higher $\mu$.
This line extrapolates to precisely the value predicted by the NJL
model at zero temperature, displayed as a large point in the figure.

We plan to continue studying the phase
diagram of this model. The dependence of the location of the tricritical
point on the mass parameter \( M \) remains to be worked out.
The addition of a non-zero bare quark mass is expected to
change the tricritical
point to a second order critical point and shift its location slightly
in the \( T-\mu  \) plane. 

\end{document}